\documentclass[conference]{IEEEtran}
\usepackage{cite}
\usepackage{amsmath,amssymb,amsfonts}
\usepackage{algorithmic}
\usepackage{graphicx}
\usepackage{textcomp}
\usepackage{xcolor}
\def\BibTeX{{\rm B\kern-.05em{\sc i\kern-.025em b}\kern-.08em
    T\kern-.1667em\lower.7ex\hbox{E}\kern-.125emX}}
\begin{document}

\title{Channel Estimation Based on Machine Learning Paradigm for Spatial Modulation OFDM}

\author{
    \IEEEauthorblockN{Ahmed M. Badi, Taissir Y. Elganimi, Osama A. S. Alkishriwo, and Nadia Adem}
    \IEEEauthorblockA{\textit{Department of Electrical and Electronic Engineering} \\
    \textit{University of Tripoli}\\
    Tripoli, Libya \\
    \{ahm.badi, t.elganimi, o.alkishriwo, and n.adem\}@uot.edu.ly}
}

\maketitle

\begin{abstract}

In this paper, deep neural network (DNN) is integrated with spatial modulation-orthogonal frequency division multiplexing (SM-OFDM) technique for end-to-end data detection over Rayleigh fading channel. This proposed system directly demodulates the received symbols, leaving the channel estimation done only implicitly. Furthermore, an ensemble network is also proposed for this system. Simulation results show that the proposed DNN detection scheme has a significant advantage over classical methods when the pilot overhead and cyclic prefix (CP) are reduced, owing to its ability to learn and adjust to complicated channel conditions. Finally, the ensemble network is shown to improve the generalization of the proposed scheme, while also showing a slight improvement in its performance.

\end{abstract}


\section{Introduction}
Spatial modulation (SM) is a promising modulation scheme at the cutting edge of wireless communication, that can achieve high spectrum and energy efficiency while remaining at a relatively low complexity level \cite{Mesleh2008}. This modulation scheme allows for the exploitation of the unique channel characteristics that the multiple transmit antennas will have, to essentially recognize the antenna’s index and utilize it as a part of the constellation diagram. On the other hand, orthogonal frequency division multiplexing (OFDM) has been considered as a well-established and a highly popular modulation scheme that enables efficient use of the available bandwidth, and to combat the inter-symbol interference (ISI) that is present in frequency selective channels. Combining the two schemes together is known as spatial modulation-orthogonal frequency division multiplexing (SM-OFDM) technique that is proposed in \cite{Mesleh2008}, where extra information bits are conveyed by active antenna indices of subcarriers to achieve enhanced data rate and robustness against the inter-antenna interference (IAI) within the subcarriers.

In SM detection schemes, the channel needs to be estimated. This has typically been accomplished using conventional methods such as least square error (LSE) and minimum mean square error (MMSE). LSE requires no prior information regarding the channel, rendering it rather simple to implement. On the other hand, the MMSE method requires second order statistics of the channel, which makes it significantly more complex. The increased complexity in MMSE is not without reason, as it improves the performance by a significant margin.

Recently, deep neural networks (DNNs) have had major development in the field of wireless communications, finding their way into many applications. DNNs specifically, have been shown promising results in channel estimation, due to their ability to learn and adapt to much more complex systems than a regular artificial neural network. In \cite{Ye2017}, the authors proposed a DNN in an OFDM system for joint channel estimation and symbol detection. The results show that the proposed DNN performs comparably to the much more complex MMSE method, and far exceeds it in low overhead situations, such as reductions in the number of pilots and the length of the cyclic prefix (CP).

\begin{figure*}[ht]
  \begin{minipage}[b]{1\linewidth}
    \centering
    \includegraphics[trim= 2.2cm 6.5cm 2.2cm 6.5cm, clip, width=0.9\textwidth]{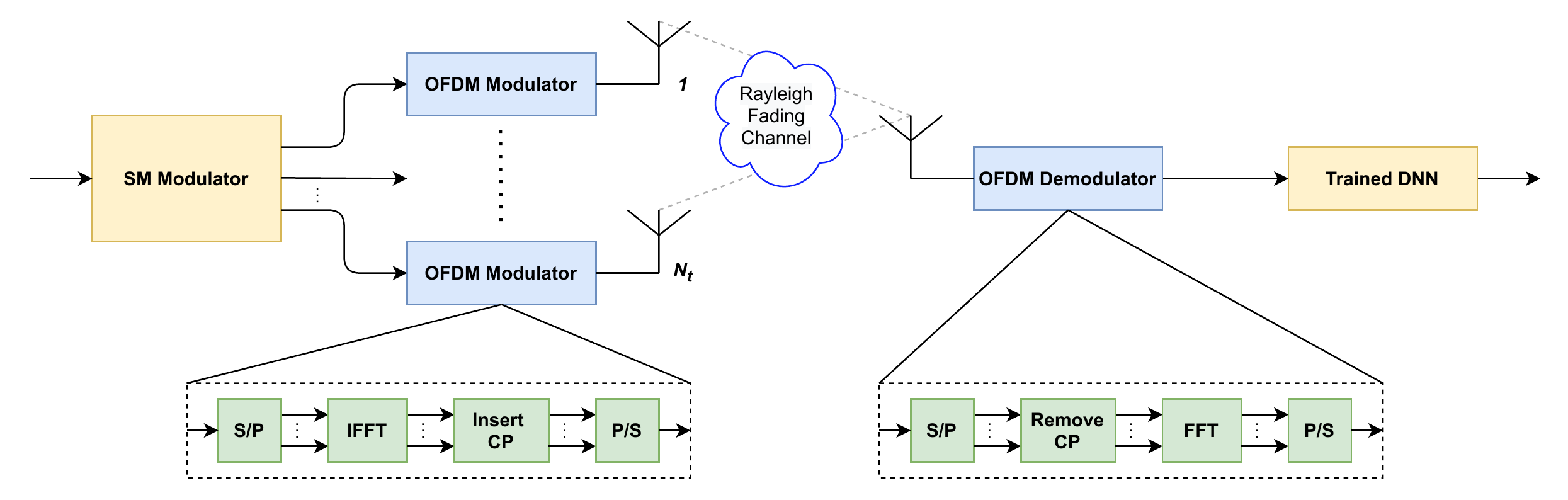}
  \end{minipage}
  \caption{SM-OFDM system model with DNN based detector.}
  \label{fig1}
\end{figure*}

The emerging SM technique is not an exception, and has also had many advancements aided by DNNs for data detection \cite{Xiang2020, Marseet2019}. For dynamic wireless channels, the authors in \cite{Xiang2020} proposed recently, a pair of DNNs to perform joint channel estimation and data detection in the SM scheme. The reported results of this paper showed that the DNN pair was capable of outperforming the conventional methods over the considered time varying channel. The authors in \cite{Marseet2019} proposed an auto-encoder complex valued convolutional neural network based detector for generalized spatial modulation (GSM) scheme, first introduced in \cite{Younis2010}, with new extracted features. It is shown in this paper that the computational complexity is significantly reduced without any noticeable performance loss as compared to the conventional maximum likelihood (ML) detector.

Ensemble learning is a popular machine learning technique, where multiple networks are merged together into a singular model, with the aspiration of improving the performance. This technique has recently been finding its way into the field of communication. For instance, in cognitive radios as in \cite{Liu2017}, where ensemble learning is used in modulation classification. It was concluded that utilizing the ensemble framework would achieve results superior to that achieved by a single classifier. Furthermore, although not explicitly stated in \cite{Ye2017}, an ensemble network was utilized, where the outputs of multiple DNNs were concatenated in order to fully detect the entire OFDM frame. This form of ensemble network allows the DNNs to be of reasonable size, such that training time does not become too egregious.

In this paper, a novel DNN scheme for joint channel estimation and data detection for SM-OFDM systems is proposed, alongside it a novel ensemble learning scheme is also proposed for the same system. This paper shows the potential of deep learning in SM-OFDM systems, given its ability to learn and adapt to frequency selective fading channels and reductions in overhead. To the best of our knowledge, this is the first attempt to utilize deep learning and ensemble learning for joint channel estimation and data detection in SM-OFDM modulation schemes.

The rest of this paper is organized as follows. Section II presents the considered SM-OFDM system architecture, the proposed DNN system, and the proposed ensemble classifier. The simulation results are presented in Section III. Finally, Section IV concludes and summarizes the paper.

\section{Proposed DNN based Detection for SM-OFDM System}
This section presents the system model of the SM-OFDM scheme, the proposed DNN based detection architecture, and the proposed ensemble network classifier.

\subsection{SM-OFDM System Architecture}
Fig. \ref{fig1} illustrates the SM-OFDM system model using DNN based detection. The SM-OFDM scheme used in this paper is similar to that developed in \cite{Mesleh2008}. At the transmitter side, the bit-stream is mapped into a matrix of size $N_t \times N_{fft}$, where  $N_t$ is the number of transmit antennas and $N_{fft}$ is the number of subcarriers. This matrix contains the quadrature amplitude modulation (QAM) or phase shift keying (PSK) symbols mapped to their respective antenna index, such that each row represents an antenna index and each column represents a subcarrier which can only be occupied by one antenna index at a time. Each row is then OFDM modulated using conventional methods. Firstly, they are transformed into a parallel signal stream, then inverse fast Fourier transform (IFFT) is applied, after that, CP is added, and finally, it is converted back to a serial signal stream to be transmitted as shown in Fig. 1.

In this paper, a sample-spaced multi-path channel is assumed. More specifically the Rayleigh fading channel model is assumed and described by $L$ complex random variables, where $L$ is the number of paths. Additive white Gaussian noise (AWGN) is also considered for this system, where only one receive antenna is equipped at the receiver, thus; the received signal can be expressed as
\begin{equation}
    y(n) = \mathbf{h}^T(n) \circledast \mathbf{x}(n) + w(n)
    \label{ch_eq}
\end{equation}
where $\circledast$ represents the time circular convolution operator, $\mathbf{h}(n)$ is the channel vector, $\mathbf{x}(n)$ is the transmitted signal vector, and $w(n)$ is the AWGN signal.

The channel is considered to remain unchanged for two consecutive SM-OFDM symbols, one of which will be mainly populated with pilots, while the other will carry the data. These two symbols together can be called a \text{frame}. For effective channel estimation, the pilot placement must be kept consistent throughout all stages of the training and deployment.

After the signal has passed through the channel and received by the receive antenna, the signal is converted to a parallel data stream so that the CP can be removed and FFT can be performed. It is then converted back to a serial stream to be fed to the proposed DNN. Thus, when compared to conventional systems, the DNN is left to perform the function of the channel estimation and the data detection of the received symbols. The DNN only outputs the recovered data, thus the channel estimation is only inferred by the artificial network.

\subsection{DNN Architecture}
\begin{figure}[ht]
  \begin{minipage}[b]{1\linewidth}
    \centering
    \includegraphics[trim= 2.54cm 15.5cm 7.92cm 2cm, clip, width=6.57cm]{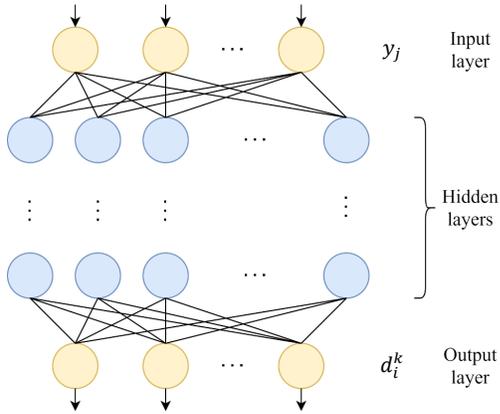}
    \vspace{-0.5cm}
  \end{minipage}
  \caption{The structure of the proposed DNN architecture for SM-OFDM signal detection.}
  \label{fig2}
\end{figure}
Due to the inherent nature of DNNs requiring a large amount of data, training needs to be performed offline, beforehand. Thus, deploying the DNN requires two stages; the first stage is offline training, and the second is online deployment of the trained network. Since the training requires a large amount of data, a channel model is expected to be employed. Channel models that have been developed by researchers are able to describe the practical channels effectively, given the exact channel statistics. Utilizing channel models allows for the simulation and generation of a large amount of data with varying channel statistics.

The proposed DNN architecture, as dipicted in Fig. \ref{fig2}, consists of six layers, where the number of neurons for these layers are 256, 1000, 500, 250, 120, and 6, respectively. The DNN takes two SM-OFDM symbols as inputs, the pilot SM-OFDM symbol and the data carrying SM-OFDM symbol. These two symbols are further separated into their real and imaginary counterparts and input into the network, thus the number of input neurons can be derived to be 256 neurons, if 64 subcarriers are used. The four hidden layers of sizes 1000, 500, 250, and 120 are activated using the \text{Relu} activation function. The output layer serves to predict and recover the received data as bits, while the activation function for this layer is the \text{Sigmoid} activation function. In our proposed system, this layer only predicts six bits, which means that a separate DNN needs to be trained for each six bits that need to be recovered. The outputs of the DNNs thus need to be concatenated in order to form an entire frame. In this way, the required size of the single DNN can be reduced significantly, while still utilizing all the data received in a single frame.

\subsection{Ensemble Network Classifier}
Multiple unique DNNs with their outputs merged together create an ensemble network. The technique of merging these networks and their quality determines the ensemble's performance and generalization. For the already proposed DNN architecture, multiple DNNs are needed to be concatenated in order to detect an entire frame, forming an ensemble network. This concatenating ensemble structure does not necessarily aim to improve the overall system’s generalization, only to reduce the size of the required DNN. Another ensemble network architecture is proposed alongside it, that aims to improve the overall system's generalization. For this proposed ensemble, the yield of multiple unique DNNs are examined, and an output is chosen based on the most probable case, according to each DNN's hypotheses. The maximum a posteriori (MAP) classifier is used, which in this particular case is equivalent to the naive Bayes optimal classifier, proposed in \cite{Mitchell1997}, due to the bit-stream being completely random. The output of the ensemble can be stated as

\begin{equation}
  d_i^{\mathrm{MAP}} = \frac{1}{K} \sum_{k=0}^{K-1} d^k_i
  \label{eq2}
\end{equation}
where $d^k_i$ is the $i$-th output of the $k$-th DNN, $d_i^{\mathrm{MAP}}$ is the ensemble’s classification for the $i$-th output, and $K$ is the number of DNNs used in the ensemble.

As \eqref{eq2} suggests, the ensemble’s classification is determined by the average of the hypotheses made by the DNNs. This is performed online, on a frame-by-frame basis. Utilizing this method allows the merging of different DNNs trained in various situations, allowing the ensemble to inherit their traits and potentially improve the overall system performance. Choosing the appropriate networks is thus vital to the ensemble’s performance and will determine its generalization.

\section{Results and Discussion}
In this section, extensive simulation is conducted to compare the bit error rate (BER) performance versus the signal-to-noise ratio (SNR) between the proposed DNN based detector and the classical LSE and MMSE detection schemes in SM-OFDM systems. Throughout the simulation, SM-OFDM scheme with two transmit antennas and one receive antenna is considered, with 64 subcarriers, a CP length of 16 samples, and modulated using the 4-QAM scheme. The Rayleigh fading channel model is also considered with the assumption that a perfect channel knowledge is available at the receiver.

\subsection{Effect of Pilots}
\begin{figure}[t]
  \begin{minipage}[b]{1\linewidth}
    \centering
    \includegraphics[trim= 0.5cm 5.5cm 0.5cm 6cm, clip, width=7.72cm]{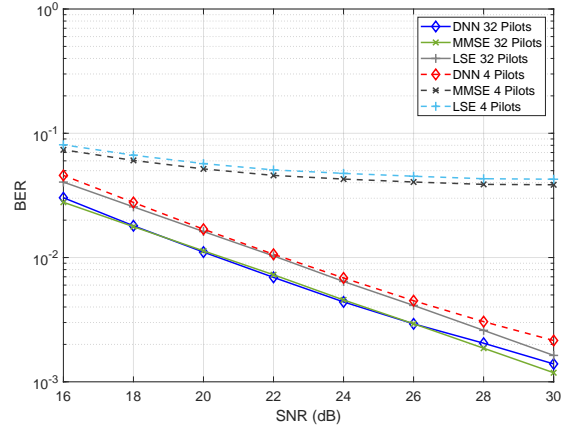}
    \vspace{-0.5cm}
  \end{minipage}
  \caption{BER performance of SM-OFDM using MMSE, LSE, and DNN based detection schemes, using 32 pilots (solid lines) and 4 pilots (dashed lines).}
  \label{fig3}
\end{figure}
In Fig. \ref{fig3}, the impact of the number of pilots on the performance of SM-OFDM scheme with the proposed DNN based detection is assessed. A Rayleigh fading channel with 3 channel paths which has a maximum delay of 3 sampling periods is considered. This figure shows that when 32 pilots per transmit antenna are used for the SM-OFDM scheme, the proposed DNN performs comparably to the conventional MMSE method and far exceeds the performance of the LSE method. Reducing the number of pilots to 4 per transmit antenna, however, significantly impacts the performance of the conventional MMSE and LSE methods, showing no significant improvement beyond the SNR of 20 dB. Meanwhile, the DNN method only suffered a performance degradation of 2 dB, as compared to the same scheme with 32 pilots to achieve a particular SNR. This comparison shows that the DNN method maintains its performance despite the large reduction in the number of pilots, making it more resilient and robust to reductions in the pilot overhead.

\subsection{Effect of Cyclic Prefix}
\begin{figure}[t]
  \begin{minipage}[b]{1\linewidth}
    \centering
    \includegraphics[trim= 0.5cm 5.5cm 0.5cm 6cm, clip, width=7.72cm]{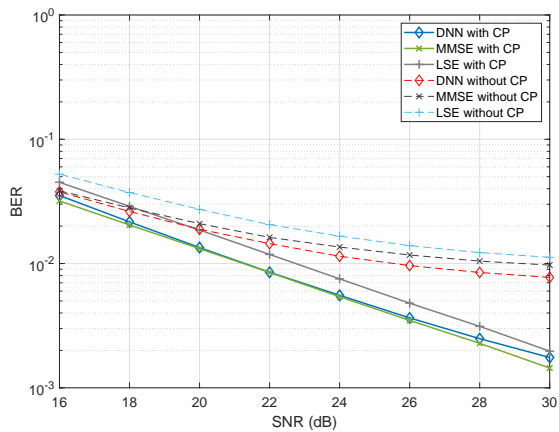}
    \vspace{-0.5cm}
  \end{minipage}
  \caption{BER performance of SM-OFDM scheme with CP (solid lines) and without CP (dashed lines).}
  \vspace{-0.05cm}
  \label{fig4}
\end{figure}
The addition of CP in OFDM systems converts the linear convolution into a circular convolution, mitigating ISI. This however, comes at the cost of wasted time and energy. In this section, the removal of the CP is evaluated. To effectively show the impact of removing the CP, a Rayleigh fading channel with 8 paths and a maximum delay of 8 sampling periods is considered. The number of pilots per transmit antenna used for all methods in this section is 32. Fig. \ref{fig4} shows that with CP, the performance is very similar to that in the previous section, with the MMSE method showing the best performance, and the proposed DNN being very comparable to it. It can also be seen from Fig. \ref{fig4} that with removing the CP, the DNN scheme outperforms the MMSE method significantly. This is due to the DNN learning and adapting to these specific conditions in ways the conventional methods cannot.

\subsection{Ensemble}
\begin{figure}[t]
  \begin{minipage}[b]{1\linewidth}
    \centering
    \includegraphics[trim= 0.5cm 5.5cm 0.5cm 6cm, clip, width=7.72cm]{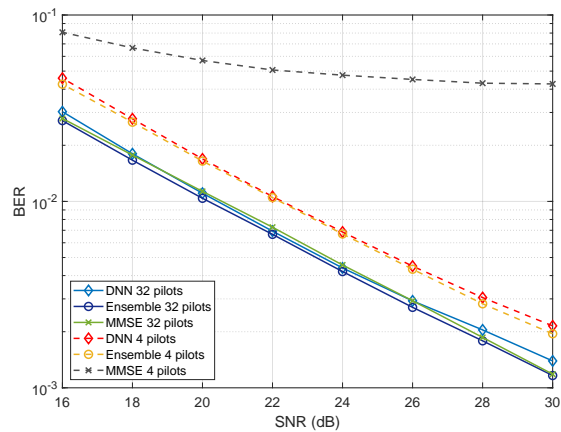}
    \vspace{-0.5cm}
  \end{minipage}
  \caption{BER performance of SM-OFDM scheme using MMSE, DNN, and ensemble schemes, using 32 pilots (solid lines) and 4 (dashed lines).}
  \label{fig5}
\end{figure}
In order to examine the effectiveness of the ensemble method, comprised of 4 unique DNNs, it is measured up against the proposed DNN and the conventional MMSE schemes. A Rayleigh fading channel with 3 channel paths and a 3 sampling period maximum delay is considered. In Figs. \ref{fig3} and \ref{fig4} it can be observed that the proposed DNN appears to be reaching BER saturation, this can be attributed to the DNNs being trained on a single SNR, leading to the DNNs to bias toward the trained SNR. The ensemble method shown in Fig. \ref{fig5}, on the other hand, does not exhibit the same level of biasing, due to the ensemble being fashioned from 4 unique DNNs trained on different SNRs. Fig. \ref{fig5} also shows that the ensemble is able to improve on the performance of the proposed DNN scheme. These results demonstrate the potential of utilizing ensembles to improve the performance of data detection based on DNNs.

\section{Conclusion and Future Work}
In this paper, a joint channel estimation and data detection architecture based on machine learning paradigm for SM-OFDM scheme is proposed. The DNN based detection scheme is evaluated and compared to the classical MMSE and LSE detection methods. Simulation results demonstrate that the DNN is able to adjust to reductions in pilot overhead and CP redundancies, while the conventional methods could not. This is due to the DNNs ability to learn and adapt to the characteristics of the wireless channel. Additionally, an ensemble network is introduced that improved the generalization of the proposed DNN architecture, while showing a slight improvement in performance. Future work will focus on further investigation of ensemble network techniques, expanding the deep learning framework into GSM, and increasing the number of transmit and receive antennas.

\bibliographystyle{IEEEtran}
\bibliography{ref.bib}

\begin{thebibliography}{1}
\providecommand{\url}[1]{#1}
\csname url@samestyle\endcsname
\providecommand{\newblock}{\relax}
\providecommand{\bibinfo}[2]{#2}
\providecommand{\BIBentrySTDinterwordspacing}{\spaceskip=0pt\relax}
\providecommand{\BIBentryALTinterwordstretchfactor}{4}
\providecommand{\BIBentryALTinterwordspacing}{\spaceskip=\fontdimen2\font plus
\BIBentryALTinterwordstretchfactor\fontdimen3\font minus
  \fontdimen4\font\relax}
\providecommand{\BIBforeignlanguage}[2]{{%
\expandafter\ifx\csname l@#1\endcsname\relax
\typeout{** WARNING: IEEEtran.bst: No hyphenation pattern has been}%
\typeout{** loaded for the language `#1'. Using the pattern for}%
\typeout{** the default language instead.}%
\else
\language=\csname l@#1\endcsname
\fi
#2}}
\providecommand{\BIBdecl}{\relax}
\BIBdecl

\bibitem{Mesleh2008}
R.~Y. Mesleh, H.~Haas, S.~Sinanovic, C.~W. Ahn, and S.~Yun, ``Spatial
  modulation,'' \emph{IEEE Trans. Veh. Technol.}, vol.~57, no.~4, pp.
  2228--2241, 2008.

\bibitem{Ye2017}
H.~Ye, G.~Y. Li, and B.-H. Juang, ``Power of deep learning for channel
  estimation and signal detection in ofdm systems,'' \emph{IEEE Wireless
  Commun. Lett.}, vol.~7, no.~1, pp. 114--117, 2017.

\bibitem{Xiang2020}
L.~Xiang, Y.~Liu, T.~Van~Luong, R.~G. Maunder, L.-L. Yang, and L.~Hanzo,
  ``Deep-learning-aided joint channel estimation and data detection for spatial
  modulation,'' \emph{IEEE Access}, vol.~8, pp. 191\,910--191\,919, 2020.

\bibitem{Marseet2019}
A.~A. Marseet and T.~Y. Elganimi, ``Fast detection based on customized complex
  valued convolutional neural network for generalized spatial modulation
  systems,'' in \emph{Proc. IEEE Western New York Image and Signal Process.
  Workshop (WNYISPW)}.\hskip 1em plus 0.5em minus 0.4em\relax IEEE, 2019, pp.
  1--5.

\bibitem{Younis2010}
A.~Younis, N.~Serafimovski, R.~Mesleh, and H.~Haas, ``Generalised spatial
  modulation,'' in \emph{Proc. Conf. Rec. 44th Asilomar Conf. Signals, Syst.
  Comput.}\hskip 1em plus 0.5em minus 0.4em\relax IEEE, 2010, pp. 1498--1502.

\bibitem{Liu2017}
T.~Liu, Y.~Guan, and Y.~Lin, ``Research on modulation recognition with ensemble
  learning,'' \emph{EURASIP J. Wireless Commun. Network}, vol. 2017, no.~1, pp.
  1--10, 2017.

\bibitem{Mitchell1997}
T.~M. Mitchell, \emph{Machine learning}.\hskip 1em plus 0.5em minus 0.4em\relax
  McGraw-hill New York, 1997.

\end{thebibliography}

\end{document}